\documentclass[prb, twocolumn, showpacs, amsmath, amssymb,superscriptaddress]{revtex4-1}
\usepackage{amssymb,amsfonts,amsmath} 
\usepackage{graphicx}
\usepackage{color}
\usepackage{hyperref}
\usepackage{longtable}
  \DeclareMathOperator{\Tr}{Tr} 
\usepackage{bbm}
\usepackage{ulem}

\newcommand{\IPR}{I\!P\!R} 
\usepackage{enumitem}
\usepackage[utf8]{inputenc}
\usepackage[T1]{fontenc}
\usepackage{lmodern}

\definecolor{darkred}{rgb}{0.75,0,0}
\begin{document} 

\title{Many-Body Localization in Two Dimensions from Projected Entangled-Pair States}

\author{D.M.\ Kennes}
\affiliation{Dahlem Center for Complex Quantum Systems and Fachbereich Physik, Freie Universit{\"a}t Berlin, 14195 Berlin, Germany}

\begin{abstract} 
Using projected entangled-pair states (PEPS) we analyze the localization properties of two-dimensional systems on a square lattice. We compare the dynamics found for three different disorder types: (i) quenched disorder, (ii) sum of two  quasi-periodic potentials along both spatial dimensions and (iii) a single quasi-periodic potential rotated with  respect to the underlying lattice by a given angle. We establish the rate of loss of information, a quantity measuring the error made while simulating the dynamics, as a good hallmark of localization physics by comparing to entanglement build-up as well as the inverse participation ratio in exactly solvable limits. We find that the disorder strength needed to localize the system increases both with the dimensionality of as well as the interaction strength in the system. The first two cases of potential (i) and (ii) behave similar, while case (iii) requires larger disorder strength to localize. 
\end{abstract}

\pacs{} 
\date{\today} 
\maketitle

\section{Introduction}
Quantum statistical mechanics relies on the ergodic hypothesis, which allows to describe systems macroscopically, i.e. independent of microscopic details. However,  in systems exhibiting many-body localization (MBL) the failure of this assumption even in generic interacting systems has recently attracted much attention.\cite{Basko2006,Gornyi2005,Huse2013,Chandran14,Imbrie2016,Nandkishore2015,Schreiber2015} As an important consequence of this failure of statistical mechanics, MBL systems do not thermalize to a statistical ensemble unraveling novel phenomena. MBL systems are now believed to be robust to small, local
perturbations and that one major aspect of MBL physics is the local preservation of information. This locality of information might harbor interesting technological applications, e.g., in the context of quantum information \cite{Huse2014,Serbyn2013,Bahri2015}.

Most theoretical studies of MBL concentrate on the 
one dimension case, which can be simulated with relative ease either using low-entanglement methods such as tensor network based approaches or exact diagonalization.\cite{Oganesyan2007,
Zelse2008,
Monthus2010,
Pal2010,
Bela2013,Kjall2014,Luitz2015,
Vasseur2015,Yevgeny2017} Under which conditions, if at all, MBL can be found in higher-dimensional systems is theoretically much more challenging. Marked exceptions of theoretical studies of MBL in 2D include a perturbatively motivated approach\cite{Yevgeny2016} as well as a quantum Monte Carlo simulations targeting excited states.\cite{Inglis2016} Intriguingly, the pioneering theoretical work of Ref.~\onlinecite{Basko2006}  suggests that  MBL does not
depend strongly on dimensionality, in contrast to other works claiming the opposite.\cite{Roeck2017,Potirniche2018,Roeck2017b}  
This renders MBL in larger than one dimension a currently hotly debated and controversial topic. From the viewpoint of quantum simulation, first experimental breakthroughs have been achieved  in controlled cold atom systems,\cite{Kondov2015,Choi2016,Bordia2017} which allow to draw some conclusions about the {\it dynamics} in 2D disordered systems prepared in a definite initial state.

Here, we connect to these experimental advances from a theory perspective using a tensor network based approach, namely projected entangled-pair states (PEPS).\cite{Orus2014,Verstraete2011} We apply this tensor network technique to a two-dimensional system on a square lattice. To connect directly to the experiments we study the transient {\it dynamics} of a system prepared initially in  a product state. Besides of the potentials realized and studied experimentally, e.g. in Refs. \onlinecite{Kondov2015,Choi2016,Bordia2017}, we will also consider a single quasi-periodic potential slanted with respect to the axis of the underlying two-dimensional system. We report a strong dependence of the localization properties on the dimensionality, interaction strength and choice of potential in general and quantify these effects.  

\begin{figure}[t]
\centering
\includegraphics[width=0.8\columnwidth]{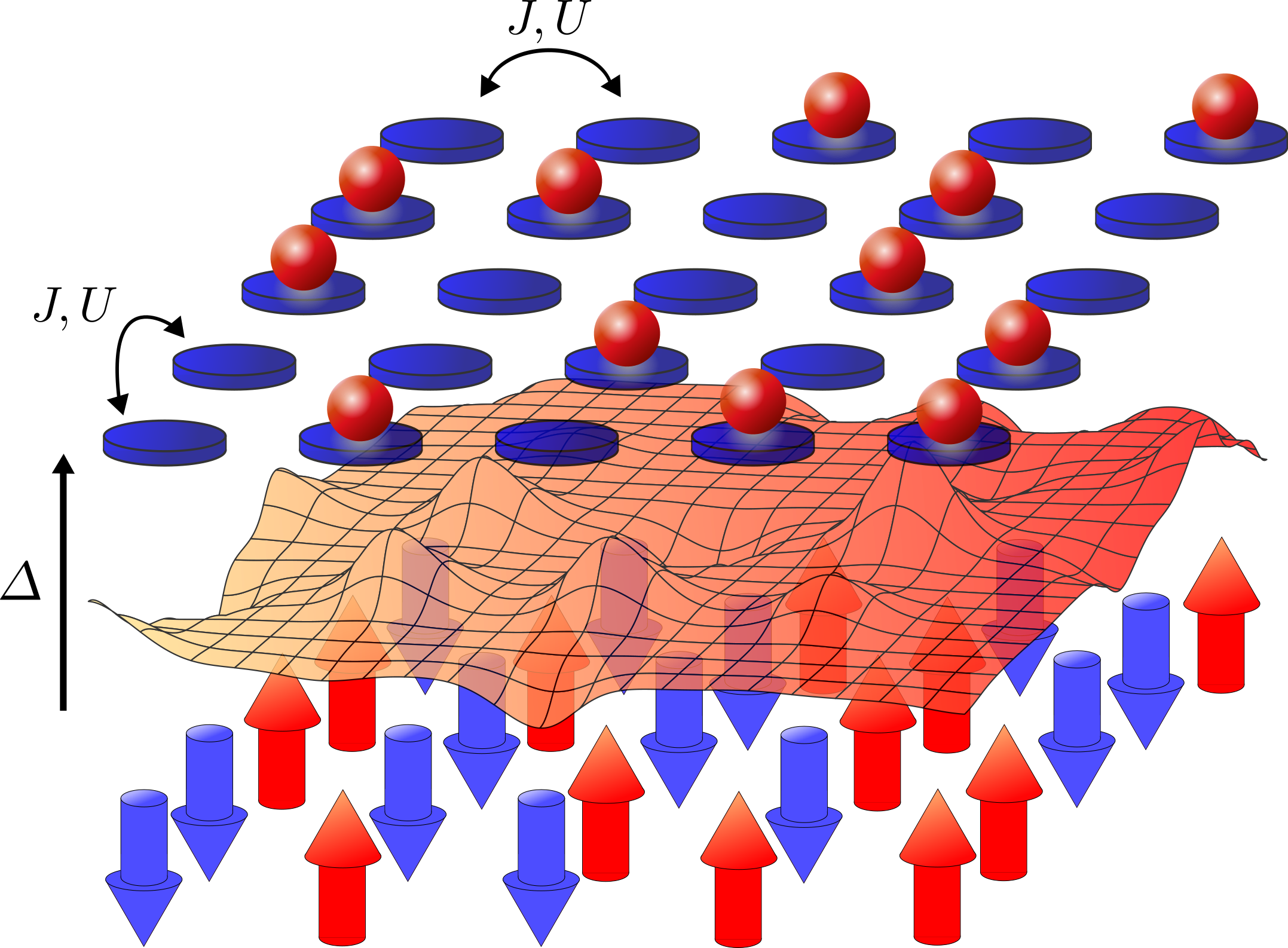}
\caption{Illustration of the studied systems. We compare spinless fermions (top) to spin-$1/2$ systems (bottom) both on a two-dimensional square lattice. We restrict the Hamiltonian to nearest neighbor terms (parametrized by $J$ and $U$). The lattice sites are subject to a random or quasi-periodic potential $\Delta$ (middle).   }
\label{fig:system}
\end{figure}

\section{Model and observables}
One of the paradigmatic models of MBL is the XXZ model 
\begin{equation}
H=J\sum\limits_{\left\langle i,j\right\rangle}\left(\sigma_x^i \sigma_x^j+\sigma_y^i \sigma_y^j\right)+U\sum\limits_{\left\langle i,j\right\rangle} \sigma_z^i \sigma_z^j+\sum\limits_{i}\Delta_i \sigma_z^i,\label{eq:Hspin}
\end{equation}
with $\sigma_{x/y/z}^i$ being the spin-1/2 Pauli matrices on site $i$. $\left\langle i,j\right\rangle$  denotes nearest neighbor on the two dimensional square lattice. We allow for $N_x$ and $N_y$ lattice sites in the x and y direction.  
We also compare to results obtained for a one-dimensional model by setting $N_y=1$.

A Jordan-Wigner transformation can be used to map the spin Hamiltonian Eq.~\eqref{eq:Hspin} to Fermions. However, in higher than one-dimension this transformation yields a long-ranged Jordan-Wigner string, which complicates calculations significantly. We here simply compare to a model omitting this string entirely:
\begin{equation}
H=J\sum\limits_{\left\langle i,j\right\rangle}\left(c_i^\dagger c_j+c_j^\dagger c_i\right)+U\sum\limits_{\left\langle i,j\right\rangle} n_i n_j+\sum\limits_{i}\Delta_i n_i.\label{eq:Hferm}
\end{equation}
Here $c_i^{(\dagger)}$ denote annihilation (creation) operators on site $i$ and $n_i=c^\dagger_ic_i$. 

The site-dependent onsite field $\Delta_i$ drives the MBL transition. We choose three different types of $\Delta_i$ (the first two of which are also motivated  experimentally\cite{Kondov2015,Choi2016,Bordia2017}).
\begin{enumerate}[label={(\roman*)}]
\item \underline{random disorder:} \\[2pt]$\Delta_i$ is drawn from a uniform distribution $\Delta_i\in [-\Delta/2,\Delta/2]$ \\(for an experimental realization see Ref.~\onlinecite{Choi2016}).
\item \underline{double quasi-periodic potential:}\\[2pt]$\Delta_i=\Delta (\cos(2\pi\beta_x i_x)+\cos(2\pi \beta_y i_y))$  \\(for an experimental realization see Ref.~\onlinecite{Bordia2017}).
\item \underline{single quasi-periodic potential:} \\[2pt]$\Delta_i=\Delta\cos\left[2\pi\beta(\vec k\cdot \vec i   )+2\pi\Theta\right]$, $\vec i =(i_x,i_y)^T$, $\vec k = (\cos(2\pi\phi),\sin(2\pi\phi))^T$ and $\Theta$ drawn from a uniform distribution $\Theta\in [0,1]$.
\end{enumerate}
The potentials (ii) and (iii) are quasi-periodic if the modulation frequency $\beta$ does not match a rational fraction of the lattice spacing. This is fulfilled for $\beta$ being (close to) an irrational number. For direct comparability we choose the experimental values of Ref.~\onlinecite{Bordia2017} $\beta_x=0.721$ and $\beta_y=0.693$ for case (ii). For case (iii) we take the inverse golden ratio $\beta=2/(1+\sqrt{5})$. The initial state is chosen as a random  product state of spins pointing up or down along $z$ corresponding to a random distribution of occupied or empty sites in the Fermionic language. 

First, we consider the average entanglement between a one or two sites cluster and the rest of the system. We denote these entanglement measured by $S_1$ and $S_2$, respectively. In the Fermionic  model and at $U=0$,\cite{Peschel2003,Peschel2009,Kennes2016} 
$
S_1(t)=-\frac{1}{D_1} \sum_i f(A_i)
$
and 
$
S_2(t)=-\frac{1}{D_2} \sum_{\left\langle i,j\right\rangle}\Tr\big[f(A_{i,j})\big],
$
with $D_i=iN_xN_y-(i-1)N_x-(i-1)N_y$ (counting the number of terms in the sums for the nearest neighbor square lattice), $f(A)=A\log(A)+(1-A)\log(1-A)$ as well as $A_i=\left\langle n_i(t) \right\rangle$ and 
\begin{equation}
A_{ij}=
\begin{pmatrix}
\langle c_i^\dagger(t) c_i(t)\rangle & \langle c_j^\dagger(t) c_i(t)\rangle\\
\langle c_i^\dagger(t) c_j(t)\rangle&\langle c_j^\dagger(t) c_j(t)\rangle
\end{pmatrix}
. 
\end{equation} 
The entanglement of a single site or a cluster of two adjacent sites with the rest of the system is bounded by $S_{i,{\rm max}}=i\log(2)$.

Secondly, we compare to the mean inverse participation ratio for the non-interacting case
\begin{equation}
\IPR_M=\frac{1}{D_1}\sum_{n=1}^{D_1}\sum_i |\phi_n(i)|^4,
\end{equation} 
with the $n$-th single particle eigenstates $\phi_n(i)$.
 
Finally, for the interacting spin model we define the so-called {\it rate of information loss} $F(t)$, which quantifies the error of simulation in tensor-network based methods like PEPS.\cite{Orus2014,Verstraete2011} When describing the dynamics of a quantum system using tensor network based methods one reformulates the quantum many-body problem by introducing tensor product states such that the wave-function can be written as $\left|\Psi^{\{\sigma_i\}}\right\rangle=\sum_{\pmb\sigma} T^{\sigma_i}\left|{\pmb\sigma}\right\rangle$, where internal contraction over the auxiliary indices $(x_1^i,y_1^i,x_2^i,y_2^i)$ of the tensor $T^{\sigma_i}_{x_1^i,y_1^i,x_2^i,y_2^i}$ are implied. If the state is lowly entangled then the dimensions of the auxiliary indeces are small, while for a highly entangled state they are (exponentially) large. In an approximate treatment using tensor networks, such as our PEPS\cite{Orus2014,Verstraete2011}, the so-called bond dimension $D^i_{x_{1/2}/y_{1/2}}$ is used to cut down the dimensions of  auxiliary indices, which is an approximation controlled by increasing $D$.\footnote{Note that in contrast to the one-dimensional case the true error of the simulation is not strictly bounded by the truncation error defined above due to the lack of left and right normalization\cite{Orus2014,Verstraete2011}} When truncating the PEPS we artificially approximate the true wave function by one with lower entanglement.\cite{Orus2014,Verstraete2011} In this sense by truncation we lose information about the (highly-entangled) portion of the wave function under scrutiny.  Our initial state  is a product state, which can be encoded with $D=1$ as its entanglement is zero. Subjected to a time evolution the bond dimension needs to grow (typically exponentially) in simulation time to keep the error of truncation $\epsilon$ below a given threshold. However, we will take a simpler vantage point using constant bond dimension during the time evolution and recording the summed truncation on all the different bonds and all times.  If the system becomes many-body localized this truncation error should drastically decrease; an intuition well established for one-dimensional systems. This picture can easily be understood on the level of entanglement scaling of the eigenstates: while for non-MBL systems the ground state usually fulfills an area law, the excited states' entanglement prototypically scales with a volume law. This leads to fast entanglement build-up for dynamical systems as excited states are scrambled into the wave function over time. In an MBL system the situation is drastically different and all Eigenstates fulfill an area-law entanglement scaling allowing to simulate dynamics very efficiently in one dimension. Moreover, the strength of localization sets the prefactor of entanglement growth with larger disorder considerable suppressing it. When the localization length become of the order of a single lattice spacing, entanglement becomes extremely short range and low bond-dimensions suffice to simulate the system accurately. With these simple considerations it appears plausible (and is tested explicitly in the following) that the localization properties can be probed by considering the rate of information loss defined via the summed truncation  error $\epsilon(t)$ as
$
F(t)=\frac{d\epsilon(t)}{dt}.
$
At small and fixed $D$, $F(t)$ should decrease rapidly at disorder $\Delta$ where the localization length becomes of the order of only a few lattice sites. This point should shift to smaller disorder as one increases $D$, as at exponentially large $D$ no error is made even at $\Delta=0$. In practice we are restricted to very small $D$ for two-dimensional systems thus we can only reliably simulate the regime where the localization length and  $S_i$ become very small and the $\IPR_M$ tends to unity. 
 
\begin{figure}[t]
\centering
\includegraphics[width=\columnwidth]{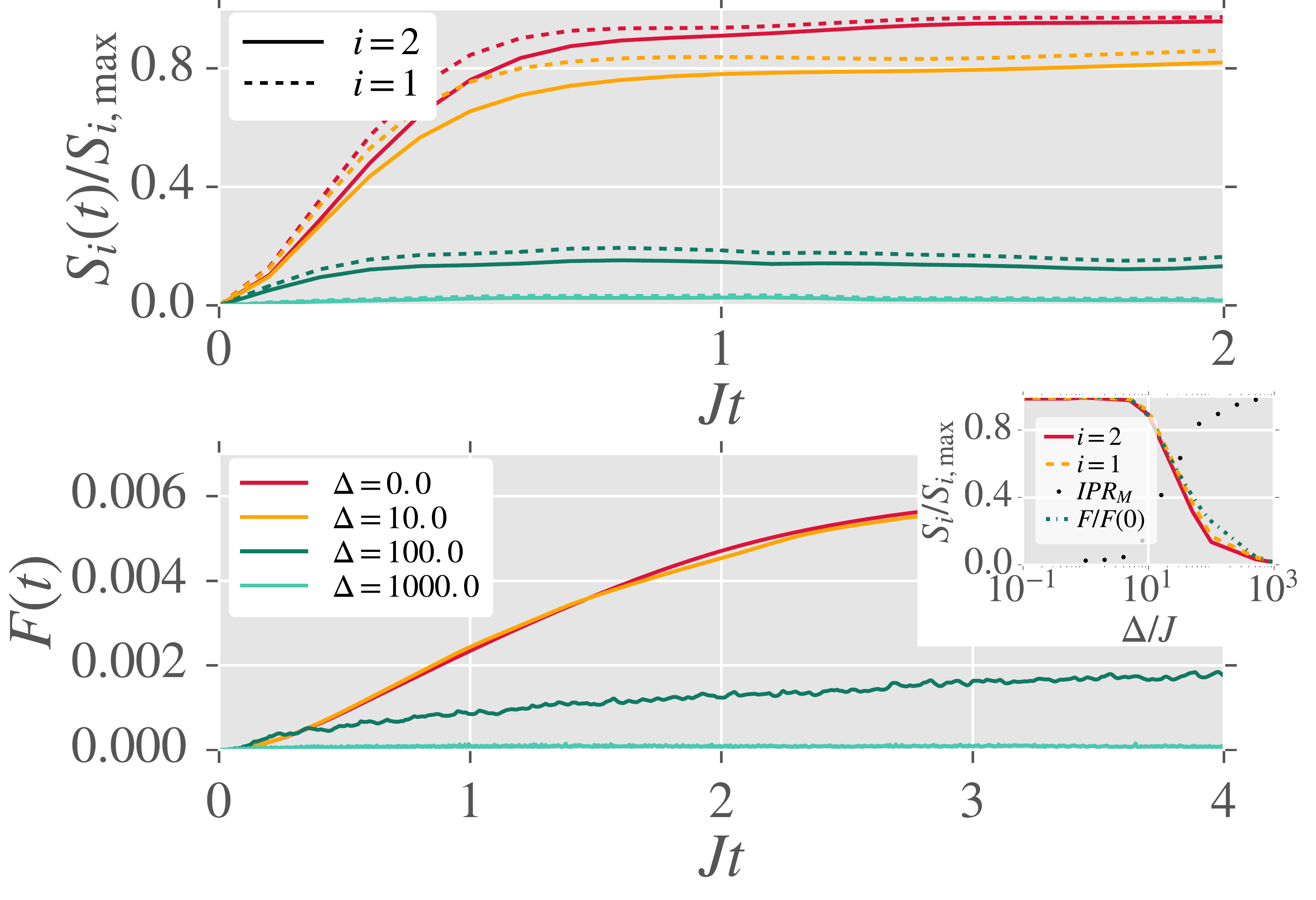}
\caption{Top: dynamical build-up of short-range entanglement, $S_1(t)$ and $S_2(t)$ for one and two sites respectively, starting from a product state in the Fermionic model. On short time scales the entanglement reaches its maximal value for $S_{1/2}$ without disorder. At large disorder the entanglement build-up is frozen. Bottom: rate of information loss $F(t)$ when truncating to a PEPS with $D=4$ and starting from a product state in the spin model. The dynamics mirror the entanglement build up of the Fermionic model. Inset: the steady value of the local entanglement $S_i=\lim_{t\to\infty}S_i(t)$ detects localization on the same scale as the mean of the inverse participation ratio $I\!P\!R_M$ and $F$. The other parameters are $N_x=N_y=10$, $U=0$ and $\Delta_{i,j}$ drawn from a random distribution of width $\Delta$.  }
\label{fig:MBL_t}
\end{figure}

\section{Results}
First, we confirm the above arguments which is exemplify in Fig.~\ref{fig:MBL_t}. The upper panel shows the average entanglement between a single or a cluster of two adjacent sites, $S_1$ and $S_2$, respectively. Initially ($t=0$) we start from a product states where the entanglement is zero. For zero disorder strength $\Delta=0$ the entanglement quickly reaches its maximum value $S_{i,{\rm max}}$. However, increasing the disorder strength the dynamics drastically change, reaching asymptotically a  much lower entangled state. The asymptotic entropy $S_i=\lim_{t\to\infty}S_i(t)$ is compared to the $\IPR_M$ in the inset. Clearly the localization measured by the $\IPR_M$ of the system is reflected in the entanglement behavior: where the $\IPR_M$ goes to unity the entanglement entropy plummets to zero. The crossover scale of both seem to be consistent, rendering both a good measure for localization. the lower panel indicates how this can be related to the rate of information loss $F(t)$ at fixed $D$. Clearly the same behavior is found. When the localization length begins to drop so does the rate of information loss. In the inset the rate of information loss at the largest accessed time $Jt=5$ is compared to the entanglement  $S_i$ and the inverse participation ratio $\IPR_M$. The rate of information loss closely mimics the behavior of the ladder two. This establishes the similarity of these quantities being either evaluated for the Fermionic model (entanglement and inverse participation ratio) or in the spin model (rate of loss of information). It also  motivates us to analyze the rate of loss of information to address the $U> 0$ case.

\begin{figure}[t]
\centering
\includegraphics[width=\columnwidth]{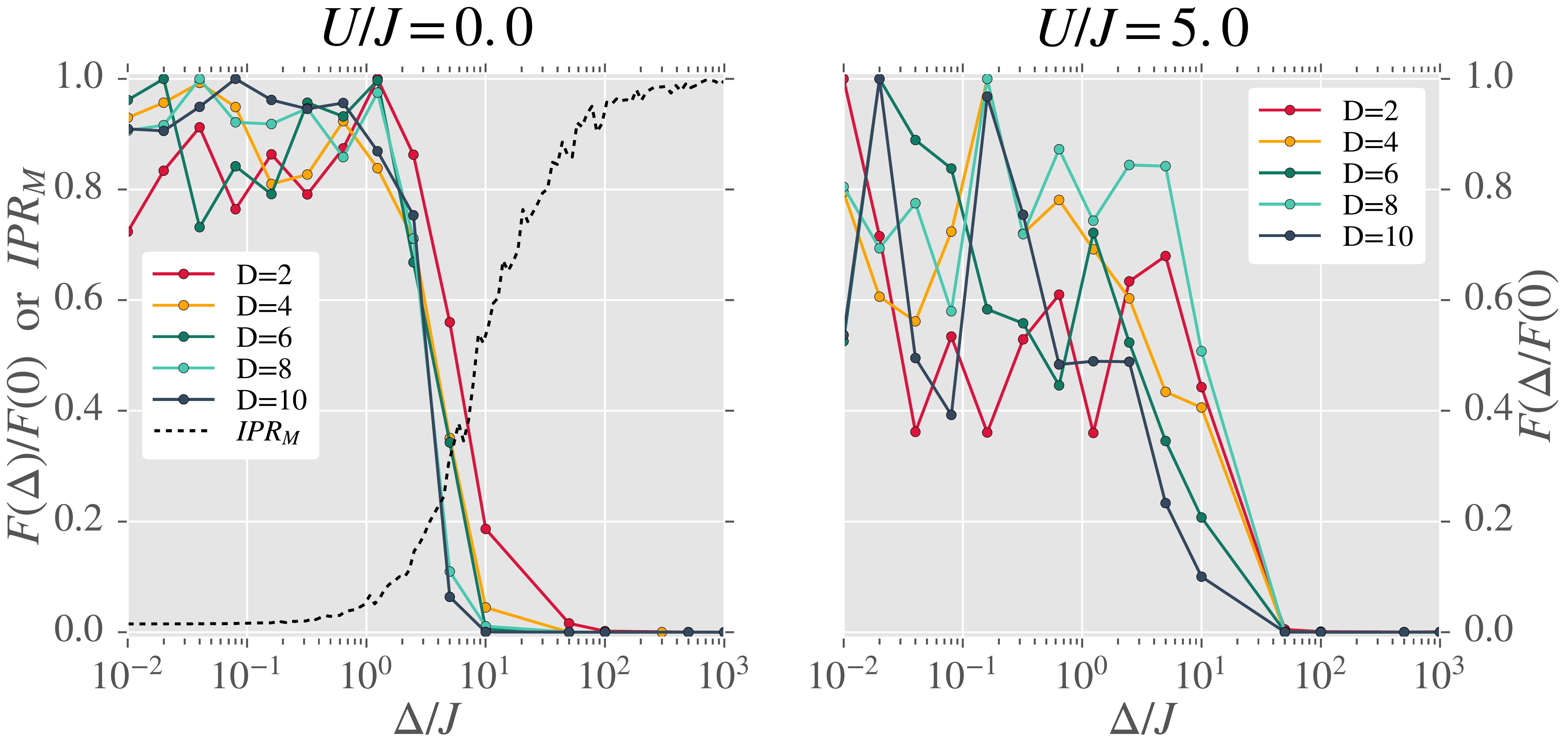}
\includegraphics[width=\columnwidth]{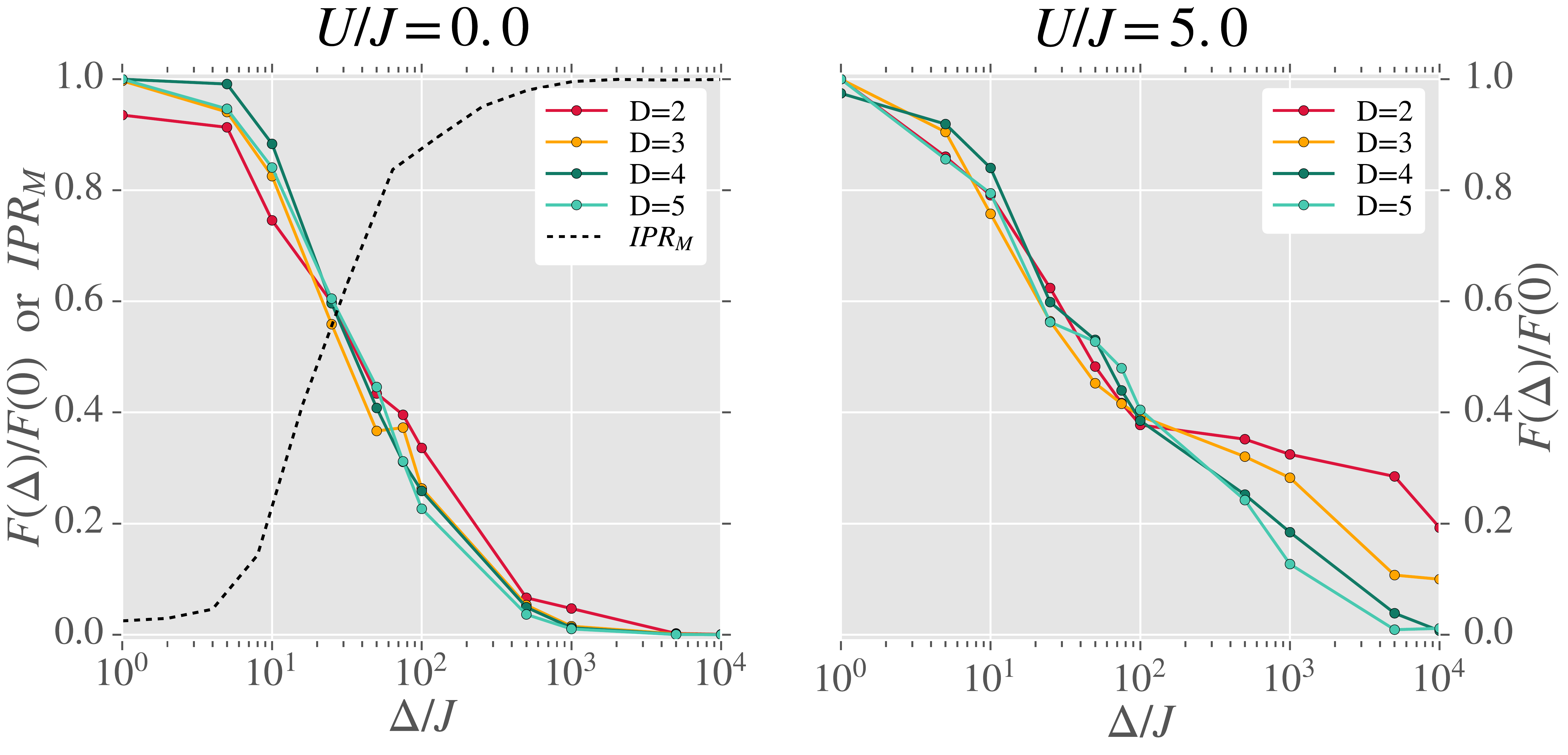}
\includegraphics[width=\columnwidth]{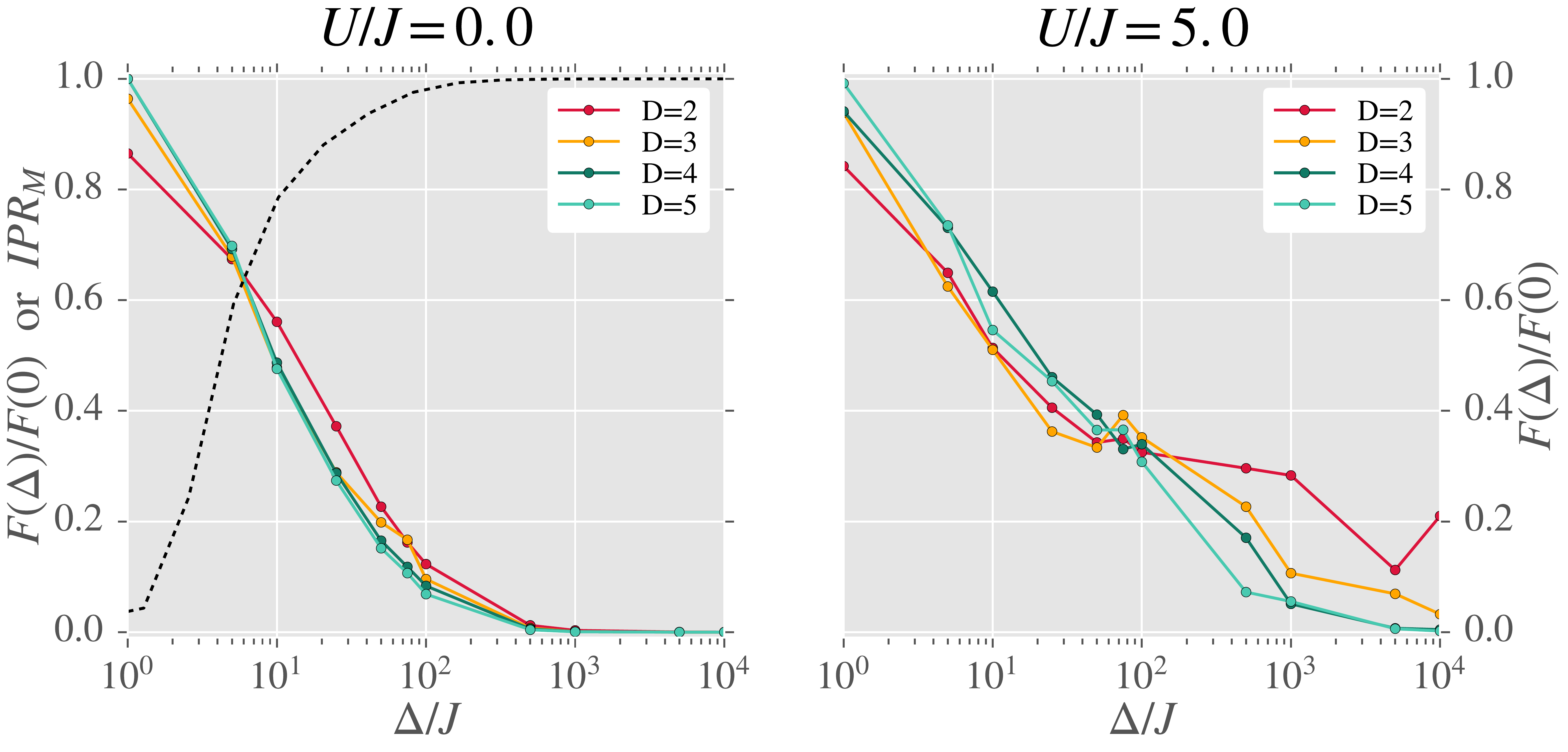}

\caption{Asymptotic rate of information loss $F$ in the spin model for random disorder in one dimension (top), random disorder in two dimensions (middle) and double quasi-periodic disorder in two dimensions (bottom). The mean inverse participation ratio $I\!P\!R_M$ for the corresponding Fermionic model is shown as a dashed line in the non-interacting limit (left column). Increasing the dimension and/or including interacting strongly delocalizes the system. The other parameters are $N_x=100$, $N_y=1$ (top panel) and $N_x=N_y=10$ (middle and bottom panel). }
\label{fig:MBL}
\end{figure}

\begin{figure}[t]
\centering
\includegraphics[width=\columnwidth]{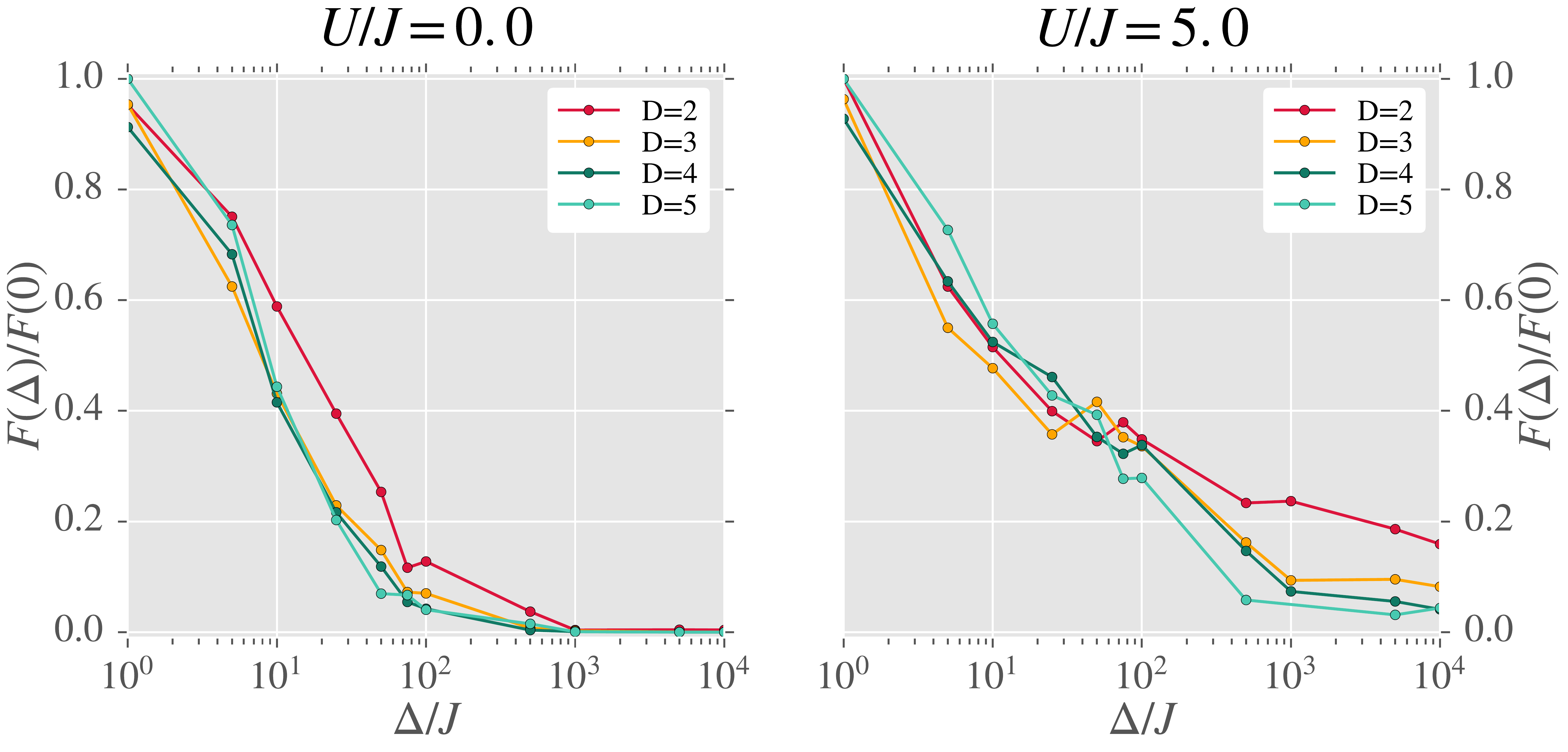}
\includegraphics[width=\columnwidth]{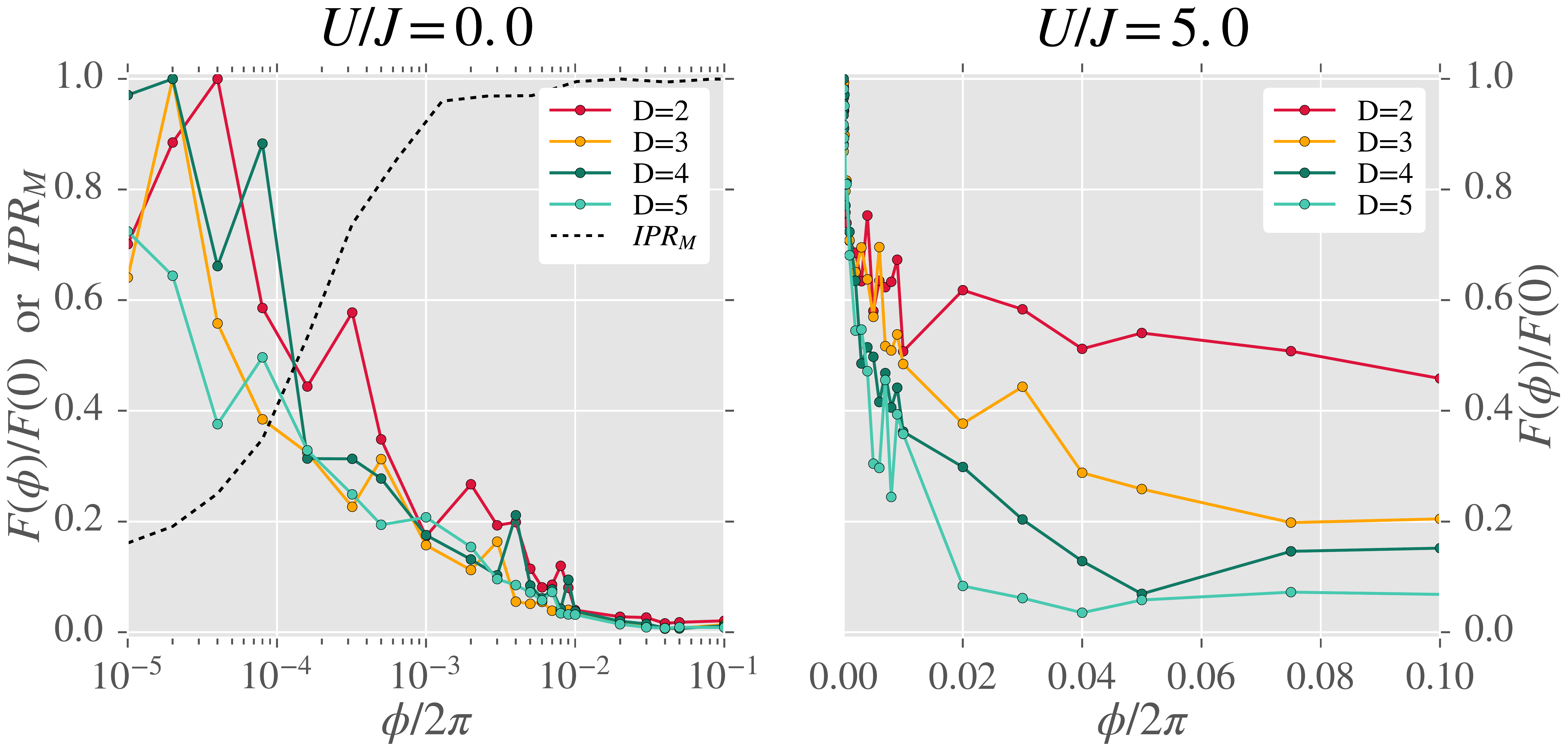}
\caption{Asymptotic rate of information loss $F$ in the spin model for a single quasi-periodic potential. Top panels: varying $\Delta$ at fixed $\phi/2/\pi=0.09$. Bottom panels: varying $\phi$ at fixed $\Delta/J=10^3$. The mean inverse participation ratio $I\!P\!R_M$ for the corresponding Fermionic model is shown for reference in the lower left panel. The other parameters are $N_x=N_y=10$.}
\label{fig:MBL_spec}
\end{figure}

 In the following, we concentrate on the rate of loss of information $F$ at the largest obtained time $Jt=5$. 
 We always compare these results in the non-interacting case $U=0$ to the mean inverse participation ratio $\IPR_M$ obtained for the Fermionic model and find generic consistence between these two measures of localization. 

In Fig.~\ref{fig:MBL} we summarize the results obtained in one-dimensional and two-dimensional systems with random disorder (top and middle row, respectively) as well as those obtained for a double quasi periodic potential in two-dimensions (bottom row). Already the analysis of the one dimensional case shows a clear transition of the rate of loss of information from a large value to almost zero as disorder is increased. The transition point is slightly moved to smaller disorder strength as $D$ is increased; as it should as $\lim\limits_{D\to\infty}F\to0$ always. The comparison to the inverse participation ratio shows that the behavior of this transition at small $D$ is well captured by $F$. This hold true for all the cases reported, providing further evidence that $F$ is a meaningful quantity to consider. Including interactions of $U/J=5$ pushes the transition to about five times larger disorder strength as the interactions tend to delocalize the system. Comparing to the two-dimensional case we report a tremendous increase in disorder strength needed to trigger the same transition in $F$ (as well as in the $\IPR_M$  for $U=0$). The point where $F(\Delta)/L(0)$ reaches $0.5$ is shifted by one order of magnitude, while the point where $L$ approaches $0$ is even increased by  two. This indicates an extreme sensitivity of MBL on the dimensionality of the system, in line with Refs.~\onlinecite{Roeck2017,Roeck2017b,Potirniche2018}. Including interactions seems to trigger a similar behavior as in the one-dimensional case. The behavior for the truly disordered system and the double quasi-periodic potential is found to be akin to each other. The disorder strength of about $\Delta/J=10$ where $F$ strongly decreases is roughly consistent with the experiment of Ref.~\onlinecite{Bordia2017} (albeit spinful fermions are used there).

Finally, we turn to the single quasi-periodic potential. Here we vary either the disorder strength $\Delta$ or the angle $\phi$. At certain angles (e.g. $\phi=0$ and $\phi=\pi$, where quasi-periodicity is only along one of the lattice directions) the potential actually becomes commensurate to the lattice structure and strong upturns in the loss of information $F$ are expected  even at large disorder. This is illustrated in the lower right panel of Fig.~\ref{fig:MBL_spec}. The results are summarized in the upper and lower row of Fig.~\ref{fig:MBL_spec} for varying disorder strength $\Delta$ and angle $\phi$, respectively. In the left and right column we compare the non-interacting to the interacting case for each. For incommensurate angles (upper row), the behavior seems to be comparable to the truly disordered or the double quasi-periodic case. However, interactions seem to play an even more significant role compared to these cases. In the lower right panel, e.g., it is shown that even $\Delta/J=10^3 $ is not sufficient to reach the same level of localization as in the truly random or double quasi-periodic potential cases for any angle $\phi$.

\section{Conclusion} We have performed a tensor network (PEPS) based analysis of  the localization physics of two-dimensional systems on a square lattice. Specifically, we analyzed the quantum quench dynamics starting from  a random product state and study the build up of entanglement.  We find that the introduced rate of loss of information, which is a measure of the error made by encoding the dynamics of a given system using a tensor network of fixed accuracy, is a good quantity to gauge the localization properties of the system. This is done by comparing explicitly to the entropy build-up as well as the mean inverse participation ratio in exactly solvable limits.  

Our analysis of three different types of potentials -- (i) quenched disorder, (ii) independent quasi-periodic potential in both of the two dimensions as well as a (iii) single quasi periodic potential with fixed angle to the square lattice-- reveals a strong  dependence of the critical disorder strength needed to localize the system on both dimensionality and interaction strength. While the potentials of class (i) and (ii) behave very similar, localization is comparably hampered in the case (iii) even away from commensurate angles, where, e.g., localization occurs only along one of the directions of the lattice.

An interesting direction of future research is the combination of real-space renormalization group \cite{RSRG1,RSRG2,RSRG3,RSRG4} and tensor network ideas akin to the study presented here. A tree tensor network could be used to simulate the system based on the real-space structure of a given disorder configuration. Utilizing the fact that entanglement build up is strongest over bonds connecting (almost) resonating onsite-energies, one should be able to choose a particularly suited tree for the tensor network, possibly opening up the route to simulate the exact dynamics in these systems up to larger times. A practical implication could use a PEPS simulation like the one done here to find over which bonds the rate of loss of information is large and in a second step choose an according tree tensor network along these bonds. 

\textit{Acknowledgements.} 
Very helpful discussions with Yevgeny Bar Lev are acknowledged.
I also acknowledge funding from the Deutsche Forschungsgemeinschaft through the Emmy Noether program (KA 3360/2-1).
Simulations were performed with computing resources granted by RWTH Aachen University under projects rwth0013 and prep0010.

\bibliographystyle{apsrev4-1}
\bibliography{2DMBL}

\end{document}